\documentclass{article}
\usepackage[utf8]{inputenc}
\usepackage[T1]{fontenc}

\usepackage{graphicx}
\usepackage{amsmath, amsthm, amssymb, amsfonts}
\usepackage{mathtools}
\usepackage{physics}
\usepackage{bm}
\usepackage{xcolor}
\usepackage[labelfont=bf,labelsep=period]{caption}
\renewcommand{\figurename}{Fig.}
\usepackage{float}
\usepackage{gensymb}
\usepackage[version=4]{mhchem}
\usepackage{multirow}
\usepackage{array}
\usepackage[super,sort&compress]{natbib}
\usepackage[colorlinks=true,citecolor=blue,linkcolor=blue,urlcolor=blue]{hyperref}

\usepackage[margin=1in]{geometry}
\usepackage{setspace}
\newcommand{\HuSR}{\textit{H\"{u}-SR}}

\begin{document}

\title{Strongly entangled Quantum Spin Rings driven by H{\"u}ckel rule}

\author{Manish Kumar$^{1,4,\dagger}$, Deng-Yuan Li$^{2,\dagger,*}$, Zhangyu Yuan$^{3,\dagger}$, Ying Wang$^{2}$, Diego Soler-Polo$^{1}$, \\ Enzo Monino$^{5}$, Libor Veis$^{5}$,  Yi-Jun Wang$^{2}$, Xin-Yu Zhang$^{6}$, Can Li$^{3}$, Jinfeng Jia$^{3,7,9}$, \\Pei-Nian Liu$^{2,*}$, Pavel Jelinek$^{1,8,*}$, Shiyong Wang$^{3,9,*}$}
\date{}

\maketitle

\noindent
$^{1}$Institute of Physics, Academy of Sciences of the Czech Republic, Cukrovarnická 10, 162 00, Prague, Czech Republic\\
$^{2}$Key Laboratory of Natural Medicines, Department of Medicinal Chemistry, China Pharmaceutical University, Nanjing 211198, China\\
$^{3}$State Key Laboratory of Micro-nano Engineering Science, Key Laboratory of Artificial Structures and Quantum Control (Ministry of Education), School of Physics and Astronomy, Shanghai Jiao Tong University, Shanghai 200240, China\\
$^{4}$Department of Condensed Matter Physics, Faculty of Mathematics and Physics, Charles University, Prague 2, CZ 12116, Czech Republic\\
$^{5}$Department of Theoretical Chemistry, J. Heyrovsky Institute of Physical Chemistry, Czech Academy of Sciences, Prague 18200, Czech Republic\\
$^{6}$School of Chemistry and Molecular Engineering, East China University of Science and Technology, Shanghai 200237, China\\
$^{7}$Hefei National Laboratory, Hefei 230088, China\\
$^{8}$Regional Centre of Advanced Technologies and Materials, Czech Advanced Technology and Research Institute (CATRIN), Palacký University Olomouc, 779 00 Olomouc, Czech Republic\\
$^{9}$Tsung-Dao Lee Institute, Shanghai Jiao Tong University, Shanghai, 200240, China\\
$^{\dagger}$These authors contributed equally\\
$^{*}$Corresponding author: dengyuanli@cpu.edu.cn, liupn@cpu.edu.cn, jelinekp@fzu.cz, shiyong.wang@sjtu.edu.cn

\clearpage
\begin{abstract}
Quantum spin rings represent an intriguing platform for studying unconventional magnetic order and exotic quantum phases, and they are also promising materials for emerging quantum technologies. Conventional spin systems consist of a set of weakly interacting localized spins that are well described by the Heisenberg spin models. Here, we demonstrate that strong interactions between radical centers in macrocycles of different sizes lead to fluctuations in the total number of unpaired electrons and to non-trivial antiferromagnetic order that extends beyond the Heisenberg picture. We demonstrate that the electronic structure of these spin rings is governed by the concept of $4n/4n+2$ H{\"u}ckel (anti)aromaticity for even-membered rings, whereas odd-membered rings possess a highly degenerate frustrated magnetic ground state. The strongly coupled spin rings are experimentally realized through the on-surface synthesis of $\pi$-magnetic carbon-based macrocycles, which consist of [2]triangulene units. The close correlation between the electronic structure and the H{\"u}ckel aromaticity rule is revealed by scanning tunneling spectroscopy and multireference calculations. This work establishes a novel design principle employing the concept of H{\"u}ckel aromaticity for quantum spin macrocycles.
\end{abstract}



The quest to construct atomically precise molecular quantum spin rings with tunable quantum properties stands at the frontier of molecular materials science~\cite{Zhang2024} and also promises new materials for emerging quantum technologies~\cite{Troiani2005}. A particularly interesting aspect is the possibility to form a frustrated magnetic order with a highly degenerate ground states~\cite{Kahn1997,Schnack2010,Baker2012}. More broadly, the preparation of new molecular magnets with ring topology exhibiting non-trivial magnetic ordering expands the portfolio of current synthetic chemistry.

Recent breakthroughs in on-surface synthesis in ultrahigh vacuum (UHV) conditions~\cite{Clair2019} have enabled the fabrication of exotic carbon allotropes, including cyclo[N]carbons, through tip-induced manipulation on insulating surfaces~\cite{Kaiser2019, Albrecht2024,Sun2023,Gao2023}. These pioneering studies have demonstrated that molecular rings containing $4n+2$ $\pi$-electrons exhibit aromatic character, whereas $4n$ systems display anti-aromatic behavior, predictions rooted in H{\"u}ckel's rule that have guided chemical intuition for nearly a century.

Series of macrocyclic polyradicaloids was synthesized with antiferromagnetically coupled $\pi$-electrons in cyclic form~\cite{Lu2016,Das2016,Prajapati2021}. In terms of aromaticity, particularly interesting are polyradicaloid polyaromatic hydrocarbon (PAH) macrocycles featuring a unique annulene-within-annulene (AWA) super-ring structure~\cite{Liu2019,Villalobos2025}. Such AWA structures show complex $\pi$-aromaticity, where the inner and outer annulene rings may exhibit distinct aromatic character following H{\"u}ckel's or Baird's rule~\cite{Liu2019}.

\begin{figure}[!htb]
    \centering
    \includegraphics[width=0.85\linewidth]{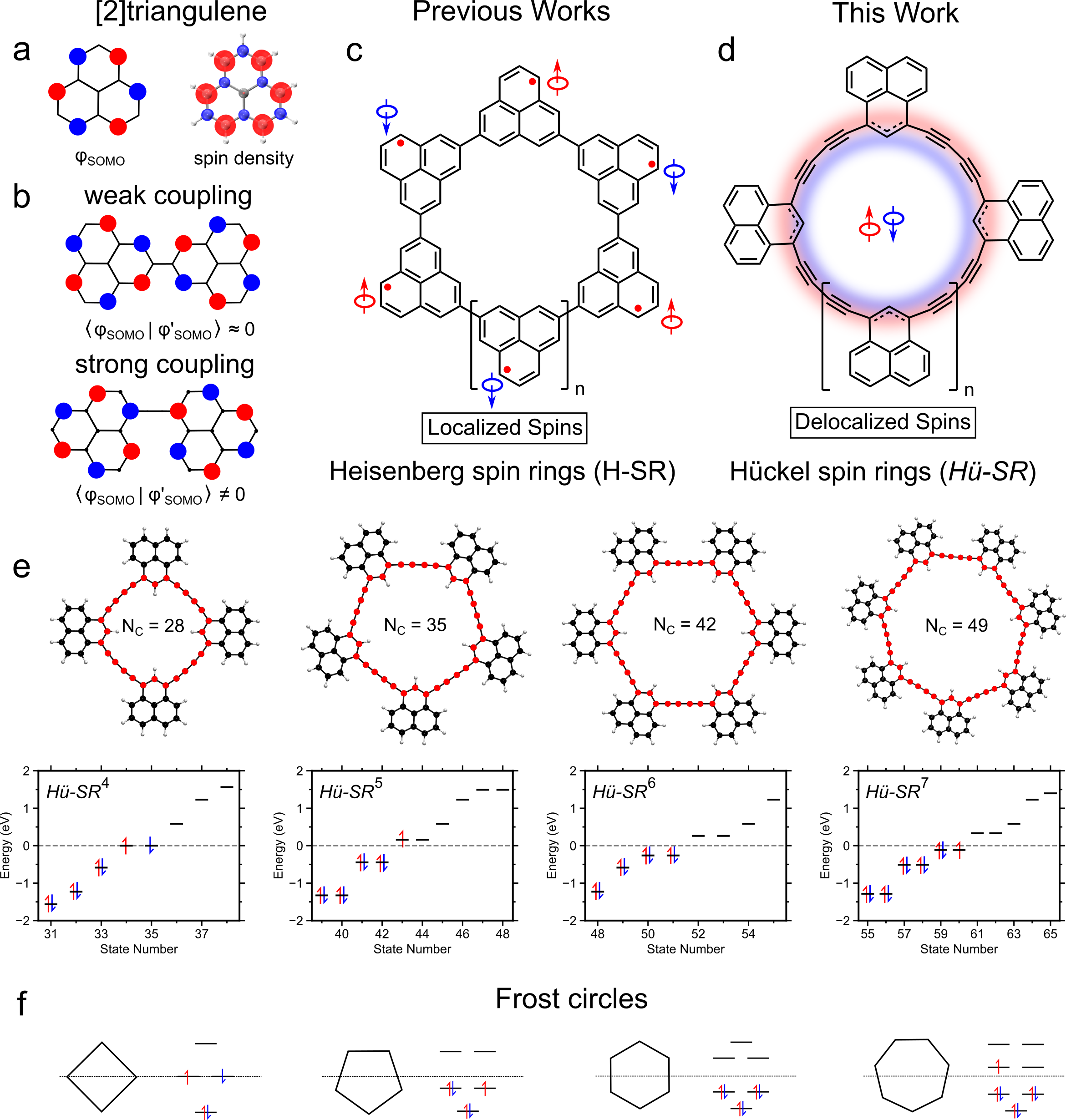}
    \caption{\textbf{Concept of H{\"u}ckel spin rings.}
\textbf{a}, Spatial localization of the zero-energy radical mode $\varphi_{\mathrm{SOMO}}$ (left) and calculated spin density of $[2]$triangulene unit.
\textbf{b}, Two connection motifs between two $[2]$triangulene units: weak coupling, where the two $\varphi_{\mathrm{SOMO}}$ do not overlap directly; and strong coupling, where the two $\varphi_{\mathrm{SOMO}}$ hybridize to form bonding and antibonding molecular orbitals.
\textbf{c}, Antiferromagnetic Heisenberg rings: localized-spin nanorings made of $[2]$triangulene units, in which the connection through specific atomic sites results in weak hybridization. Their magnetic properties follow the Heisenberg spin model and are driven by local antiferromagnetic exchange rather than by the global $\pi$-topology.
\textbf{d}, H{\"u}ckel spin rings, in which $[2]$triangulene units are connected via polyynic C$_4$ linkers. This connectivity enforces strong hybridization, resulting in delocalized spins and magnetic properties governed by H{\"u}ckel's $4n/4n+2$ rules.
\textbf{e}, Atomic structures of H{\"u}ckel spin rings with different numbers of $[2]$triangulene units and their corresponding one-electron energy spectra. Carbon atoms (highlighted in red) form an inner ring whose electron count determines the $4n$ (antiaromatic) or $4n+2$ (aromatic) character.
\textbf{f}, Frost circles for the rings in (e), rationalizing the alternation of radical character with the number of $[2]$triangulene units.}
    \label{fig:fig1}
\end{figure}

More recently, [n]triangulene units have been used as building blocks to construct low-dimensional $\pi$-magnets with unconventional electronic and magnetic properties, including linear chains~\cite{yuan2025fractional,Mishra2021,Zhao2022,Zhao2025} and rings~\cite{Hieulle2021,li2025frustration}. In these works, the presence of well-defined localized spins arises from weak hybridization between singly occupied zero-energy modes $\varphi_{\mathrm{SOMO}}$ of triangulene units originating from the imbalance of the bipartite lattice~\cite{Ovchinnikov1978}. For [2]triangulene, the presence of a~single $\varphi_{\mathrm{SOMO}}$ (Fig.~\ref{fig:fig1}a) gives rise to spin-1/2 building block. The weak coupling is caused by a particular connection through atomic sites of the bipartite lattice of [2]triangulene that do not host the radical state $\varphi_{\mathrm{SOMO}}$ (Fig.~\ref{fig:fig1}b, top). Consequently, spin-1/2 characters that originate from a topological zero energy mode with antiferromagnetic (AFM) coupling between neighbor spins are retained, as depicted in Fig.~\ref{fig:fig1}c and  Fig. S1a. Consequently, the magnetic properties reflect Heisenberg's picture, driven by local antiferromagnetic exchange interactions between adjacent well-defined spin centers rather than the global $\pi$-electron topology, thereby preventing any significant role for aromaticity.

In parallel, odd-membered $\pi$-magnetic rings, lying outside the conventional $4n/4n+2$ aromatic classifications, have attracted growing interest due to their frustrated spin topology~\cite{Schnack2010,Kahn1997}. For odd-membered AFM spin-1/2 rings, a fourfold-degenerate ground state is composed of two symmetry-related doublets. Their degenerate electronic structure provides a complementary platform where geometric frustration, rather than H{\"u}ckel aromaticity alone, defines the strongly correlated ground state. In a recent study of five-membered triangulene rings~\cite{li2025frustration,Li_Arxiv},
steric hindrance between adjacent triangulene units, imposed by a short linker, prevented the formation of the frustrated ground state. Consequently, frustrated odd-membered $\pi$-magnetic rings have so far remained largely theoretical, and an experimental platform that clearly manifests their magnetic frustrated ground states is still lacking.

Here, we introduce a design strategy that employs the concept of H{\"u}ckel rules in quantum spin rings. We consider extended $\pi$-conjugated macrocycles in which spin-1/2 [2]triangulene units are specifically connected through atomic sites that host the singly occupied radical state $\varphi_{\mathrm{SOMO}}$ (Fig.~\ref{fig:fig1}b, bottom) via a diyne C$_4$ linker. This connectivity enforces strong hybridization between adjacent radical states, i.e., $\langle \varphi_{\mathrm{SOMO}}|\varphi'_{\mathrm{SOMO}}\rangle \neq 0$, and they hybridize to form bonding and antibonding molecular orbitals (Fig. S1c). In the dimer, strong hybridization together with electron-electron correlation leads to the open-shell singlet ground state~\cite{Turco2025} as shown experimentally and theoretically in Fig. S1-3. Extending this motif to rings, the strong hybridization of the radical states $\varphi_{\mathrm{SOMO}}$ within the cycle leads to delocalization of the electronic states and produces antiferromagnetic order over the entire ring (Fig.~\ref{fig:fig1}c). For convenience, we refer to these molecular spin rings as the H\"{u}ckel- spin ring (\HuSR$^N$), hereinafter, where $N$ corresponds to the number of [2]triangulene units for reasons discussed later.

The effect of the delocalization reflected in the evolution of the character of one-electron H{\"u}ckel spectra of different \HuSR$^N$, as shown in Fig.~\ref{fig:fig1}e. For \HuSR$^4$, 28 carbon atoms (highlighted in red in Fig.~\ref{fig:fig1}e) form an inner carbon cycle corresponding to $4n$ $\pi$-electron count and thus to anti-aromatic character. Consequently, according to H{\"u}ckel's rules, there are two topological zero-energy modes $\varphi_{\mathrm{SOMO}}$ hosting two unpaired electrons (Fig.~\ref{fig:fig1}e), giving \HuSR$^4$ a strong diradical character. In contrast, \HuSR$^6$ has $N_c=42$ carbon atoms in the inner cycle, which satisfies the $4n+2$ aromatic rule; its one-electron spectrum therefore shows closed-shell character, albeit with a relatively small band gap (Fig.~\ref{fig:fig1}e).
 
For odd-membered \HuSR, the H{\"u}ckel energy spectrum exhibits doubly degenerate frontier orbitals $\varphi_{\mathrm{SOMO}}$ that host one (\HuSR$^5$) or three (\HuSR$^7$) electrons, respectively (Figs.~\ref{fig:fig1}e). For completeness,  Fig. S4 shows the one-electron spectra for larger rings \HuSR$^{8-12}$, which reveal alternating $4n/4n+2$ character for even-membered \HuSR depending on the number of carbon atoms ($N_c$) in the inner ring. For odd-membered rings, two doubly degenerate frontier orbitals are always present and host one or three unpaired electrons. Overall, this analysis indicates that for $\pi$-magnetic \HuSR\ with strongly hybridized radical states $\varphi_{\mathrm{SOMO}}$, the electronic structure follows the Frost-circle picture (Fig.~\ref{fig:fig1}f) imposed by the cyclic boundary conditions. However, electron correlation plays a central role in \HuSR, so the one-electron H{\"u}ckel model and Frost circles provide radical originating only from the topological zero energy modes, and multireference methods are required for a quantitative description, as discussed below. 

\begin{figure}[!htb]
    \centering
    \includegraphics[width=0.9\linewidth]{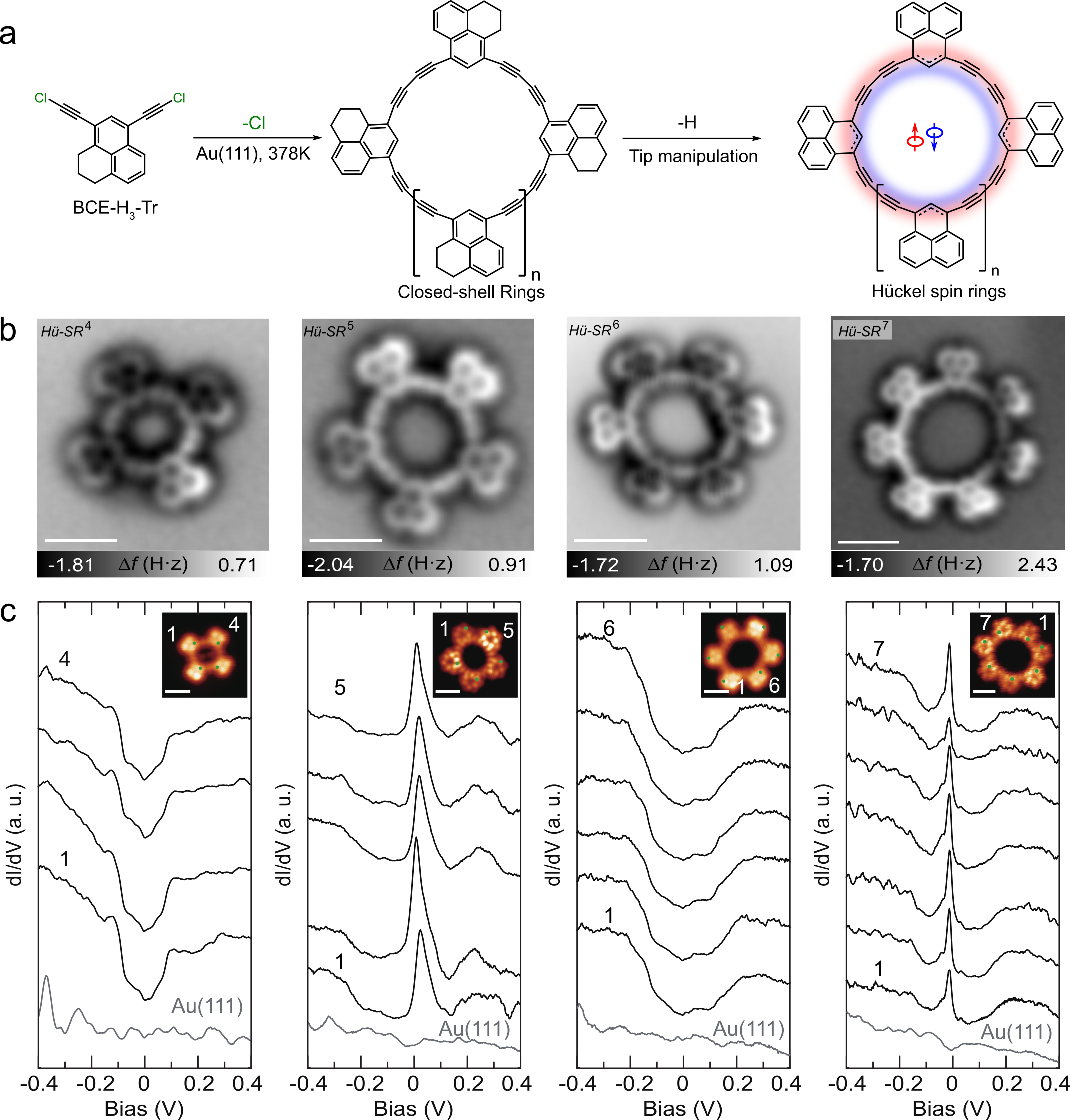}
    \caption{\textbf{Design and synthesis of molecular H{\"u}ckel spin rings.}
\textbf{a}, Schematic illustration of the synthetic pathway, involving surface-assisted dechlorination C--C coupling followed by STM tip-induced dehydrogenation.
\textbf{b}, AFM images of molecular spin rings composed of alternating [2]triangulene units and diyne bridges. Scale bar: 1 nm.
\textbf{c}, STS ($dI/dV$) spectra of macrocycles \HuSR$^4$ to \HuSR$^7$. Even-membered rings (\HuSR$^4$ and \HuSR$^6$) display symmetric step-like features characteristic of inelastic spin-flip excitations, whereas odd-membered rings (\HuSR$^5$ and \HuSR$^7$) exhibit zero-bias resonances. Scale bar: 1 nm.}
    \label{fig:on_surface_synthesis}
\end{figure}

To corroborate these results, an on-surface synthesis approach was employed to prepare the desired \HuSR. To achieve the construction of molecular spin rings with even and odd numbers [2]triangulene units, we designed and synthesized a molecular precursor with a closed shell structure, 4,6-bis(chloroethynyl)-2,3-dihydro-1H-[2]triangulene (BCE-H$_3$-Tr) by solution methods (see detailed synthesis and structural characterization in the Supplementary Material). On the naphthalene ring of BCE-H$_3$-Tr, two meta-chloroethynyl groups are expected to form diyne bridges on the Au(111) surface via an intermolecular dechlorination C-C coupling reaction~\cite{shu2020atomic}, and 1,3-propylene (CH$_2$CH$_2$CH$_2$) can undergo STM tip-induced dehydrogenation to achieve the transformation from sp$^3$ to sp$^2$ hybridized carbon~\cite{li2025programmable}, resulting in the formation of diyne-bridged [2]triangulene polymers. As shown schematically in Fig.~\ref{fig:on_surface_synthesis}a, the molecular precursors BCE-H$_3$-Tr were deposited onto a clean Au(111) surface held at 378 K under ultrahigh vacuum conditions. After annealing at 378 K for 1 h, STM measurements revealed the formation of the linear polymer chains and cyclic oligomers ( Fig. S5). Subsequently, STM tip-induced dehydrogenation of cyclic oligomers yielded molecular spin rings with $N=4$ to 13 [2]triangulene units (see Fig. S5d-g and Fig. S6). Bond-resolved nc-AFM images of representative \HuSR$^{4-7}$ confirm alternating [2]triangulene units and diyne bridges (Fig.~\ref{fig:on_surface_synthesis}b).

Fig.~\ref{fig:on_surface_synthesis}c shows experimental scanning tunneling spectroscopy (STS) spectra acquired for different macrocycles \HuSR$^4$ to \HuSR$^7$.  The spectra exhibit distinct low-energy excitations that depend on whether the number $N$ of [2]triangulene units is even or odd. For even-membered rings ($N=4,6$), symmetric step-like features were observed that are commonly attributed to inelastic tunneling processes associated with spin-flip transitions between the ground and excited states~\cite{Hirjibehedin2007}. For odd-membered macrocycles, we detect zero-bias resonances, consistent with Kondo screening of molecular spins on metal surfaces~\cite{Hewson1993}. For \HuSR$^7$ and larger rings, step-like features associated with spin-flip excitations are also observed. The step-like signal is not clearly resolved for \HuSR$^5$, which we attribute to the proximity of ionic resonances in the same energy range in the STS, hindering the detection of the spin excitation signal.

\begin{figure}[!htb]
    \centering
    \includegraphics[width=1\linewidth]{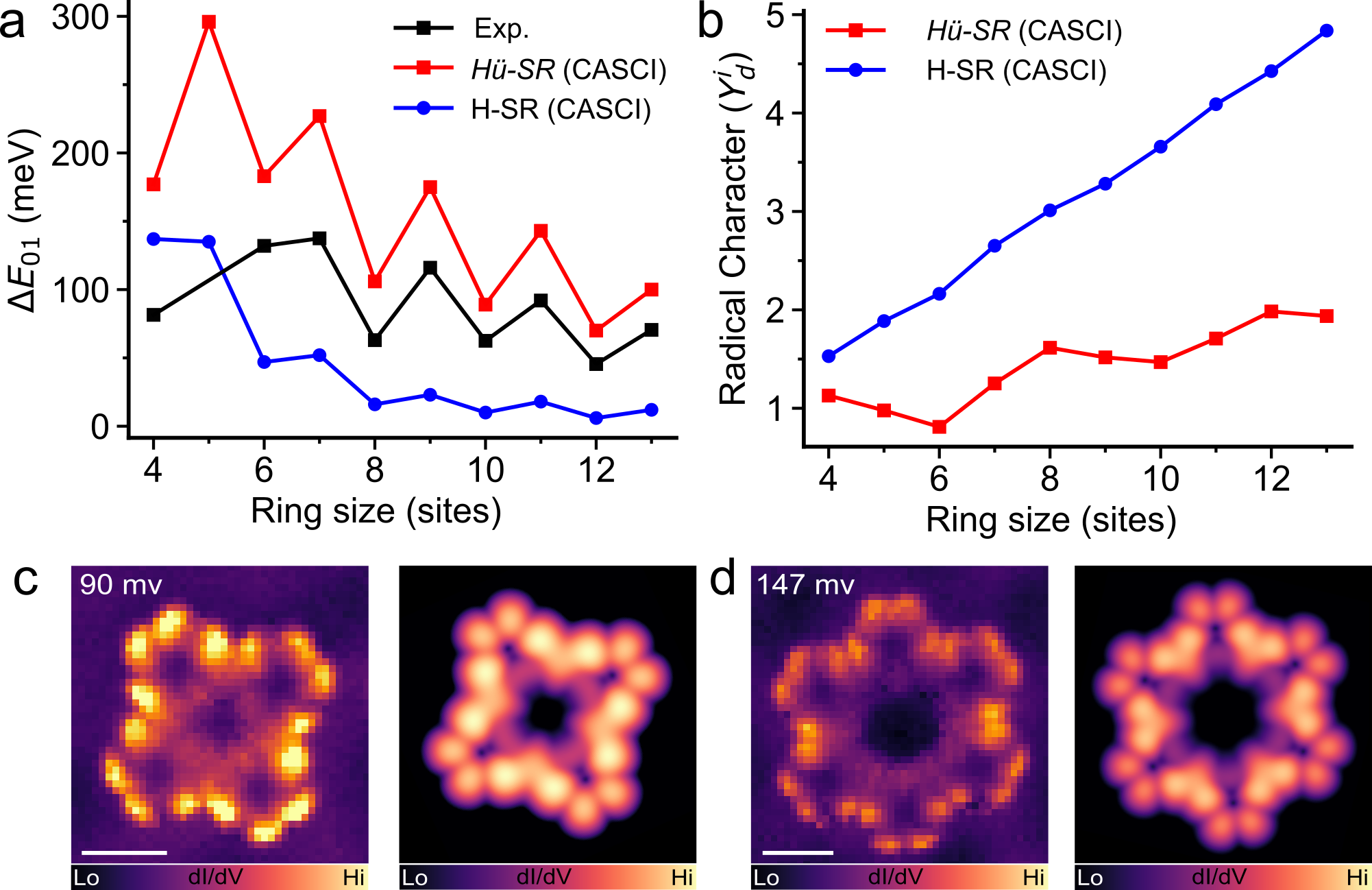}
    \caption{\textbf{Magnetic properties of molecular spin rings.}
\textbf{a}, Lowest spin-excitation energies $\Delta E_{01}$ as a function of ring size $N$ obtained from experimental STS (black), CASCI calculations for \HuSR$^N$ (red), and the Heisenberg spin ring (H-SR, blue) shown in Fig.~\ref{fig:fig1}c.
\textbf{b}, Total radical character $Y_d$ versus ring size $N$ for \HuSR\ (red) and the Heisenberg spin ring (H-SR, blue).
\textbf{c}, Comparison of experimental constant-current STS maps acquired at 90 meV with simulated STS maps computed from CASCI NTOs for the singlet--triplet excitation of \HuSR$^4$.
\textbf{d}, Comparison of experimental constant-current STS maps acquired at 147 meV with simulated STS maps computed from CASCI NTOs for the singlet--triplet excitation of \HuSR$^6$. Scale bars: 1 nm.}
    \label{fig:even_ring}
\end{figure}

Fig.~\ref{fig:even_ring}a (black square) summarizes the lowest spin-excitations $\Delta E_{01}$ obtained from experimental STS spectra for \HuSR$^N$ up to $N=13$. The spin-excitation data exhibit an even-odd oscillation and, overall, a decrease with increasing $N$. Notably, the size dependence is not strictly monotonic as $\Delta E_{01}$ for \HuSR$^4$ is lower than that of \HuSR$^6$, in contrast to the behavior expected for a Heisenberg spin ring (blue circles).

To rationalize the experimental STS measurements and better understand the electronic structure of \HuSR, we performed multireference complete active space configuration interaction (CASCI) calculations. We used geometries optimized in the gas phase with density functional theory (DFT) (see  Fig.~S5). Whereas single-determinant DFT favours a broken-symmetry solution, multireference CASCI yields a lower energy for fully symmetric geometries, indicating the absence of a Jahn-Teller distortion in \HuSR\ (see Supplementary Material).

Fig.~\ref{fig:even_ring}a compares experimental and theoretical values of the lowest spin excitation $\Delta E_{01}$ showing very good agreement. We attribute overall difference in the absolute values of the spin excitation $\Delta E_{01}$ between the experiment and the theoretical CASCI calculations to the absence of metal surface screening in the multireference calculations, which generally reduces the value of the spin excitation~\cite{zuzak2024surface}. In particular, we would like to emphasize that the experimentally observed decrease of the spin excitation $\Delta E_{01}$ between \HuSR$^6$ and \HuSR$^6$ is well reproduced by the multireference CASCI calculations.

For even-membered \HuSR, the calculated electronic structure shows the multireference open-shell singlet as ground state and triplet state as the first excited state (Supplementary Figs. S10--15). Importantly, the singlet-triplet excitation energy $\Delta E_{01}$ shows non-monotonic character with $N$, which is inherently connected to variation of (anti)aromatic character and associated one-electron H{\"u}ckel spectra of \HuSR$^N$. For anti-aromatic \HuSR$^4$, which in one-electron picture has open-shell character with two degenerate topological zero energy modes $\varphi_{SOMO}$ (Fig.~\ref{fig:fig1}e), CASCI(12,12) calculation gives $\Delta E_{01}=177$ meV. Nevertheless, for aromatic \HuSR$^6$, multireference calculations give a higher singlet-triplet excitation $\Delta E_{01}=183$ meV, in good agreement with the experimental observation. Here, the increase of $\Delta E_{01}$ is caused by the fact that H\"{u}ckel energy spectra show close shell character. However, in the multireference CASCI calculations, the enhanced electron-electron correlation facilitated by a relatively small one-electron band gap~\cite{Biswas2023,Song2024} gives rise to an open-shell singlet but with a larger $\Delta E_{01}$ compared to \HuSR$^4$. This scenario is also confirmed by the calculated radical character $Y_d^i$ proposed by Yamaguchi et al~\cite{yamaguchi1988extended} using CASCI calculations (Fig.~\ref{fig:even_ring}b, red line), which evolves non-monotonically with N number of [2]triangulene units. Namely, we observe that the radical character decreases from \HuSR$^4$ to \HuSR$^6$. In general, the radical character of \HuSR\ exhibits a significantly different behavior compared to the Heisenberg spin rings (blue line), whose radical character is generally higher and almost linear with the number $N$ of [2]triangulene units. These theoretical comparisons provide direct evidence of the role of (anti)aromatic behavior on the magnetic properties of \HuSR$^N$, which deviates from the traditional Heisenberg picture. Finally, \HuSR\ exhibits substantially larger spin-excitation energies $\Delta E_{01}$ than previously reported Heisenberg-type spin rings, reflecting the strong hybridization of the radical states.

Figs.~\ref{fig:even_ring}c,d compare experimental and simulated STS maps obtained from natural transition orbitals (NTO) corresponding to the singlet-triplet transition for \HuSR$^{4,6}$. Very good agreement between experimental and simulated STS maps confirms that the step-like feature in STS is associated with the singlet-triplet transition.

\begin{figure}[!htb]
    \centering
    \includegraphics[width=1\linewidth]{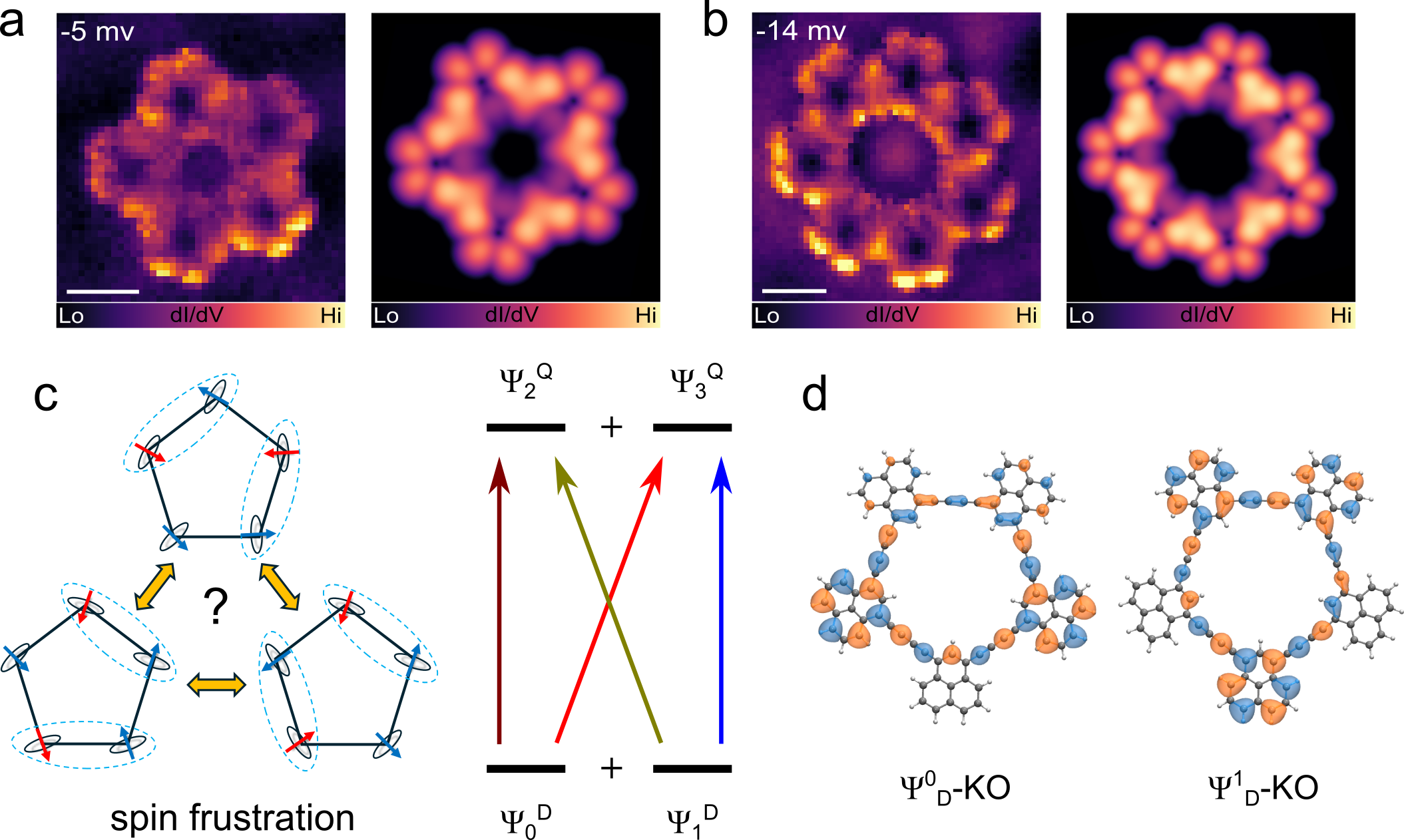}
    \caption{\textbf{Frustrated magnetic order of odd-membered \HuSR.}
\textbf{a}, Comparison of experimental constant-current STS maps acquired at $-5$ meV with simulated constant-height STS maps computed from CASCI Kondo orbitals associated with Kondo screening of the degenerate doublet ground states $\Psi^D_{0,1}$ of \HuSR$^5$ on a metal surface.
\textbf{b}, Comparison of experimental constant-current STS maps acquired at $-14$ meV with simulated constant-height STS maps computed from CASCI Kondo orbitals associated with Kondo screening of the degenerate doublet ground states $\Psi^D_{0,1}$ of \HuSR$^7$ on a metal surface.
\textbf{c}, Schematic representation of frustrated spin order in \HuSR$^5$, showing competing local magnetic configurations with entangled singlets (dashed ovals, right). Energy-level scheme for odd-membered \HuSR\: doubly degenerate doublet ground state $\Psi^D_{0,1}$ and first excited quartet states $\Psi^Q_{2,3}$, indicating magnetic frustration.
\textbf{d}, Calculated CASCI Kondo orbitals associated with screening of the two degenerate doublet states $\Psi^D_{0,1}$ by the metallic substrate, highlighting their distinct spatial distributions. Scale bars: 1 nm.}
    \label{fig:odd_ring}
\end{figure}

For odd-membered \HuSR, STS reveals the presence of zero-energy resonances tentatively associated with Kondo screening (Fig.~\ref{fig:on_surface_synthesis}c). Constant-current STS maps acquired near the Fermi level show that the zero-energy signal is distributed homogeneously over the entire \HuSR\ (Figs.~\ref{fig:odd_ring}a,b). According to CASCI(11,11) calculations, the ground state of odd-membered \HuSR\ consists of two doubly degenerate doublet states $\Psi^D_{0,1}$ with strong multireference character (see Supplementary Figs. S17--S23), arising from frustrated antiferromagnetic interactions between spin-1/2 radicals (Fig.~\ref{fig:odd_ring}). The first excited states form a manifold of doubly-degenerate quartet states $\Psi^Q_{2,3}$. 

To rationalize the zero-energy resonances in terms of Kondo screening, we computed Kondo orbitals~\cite{calvo2024theoretical} corresponding to Kondo screening of two degenerate doublet ground states $\Psi^D_{0,1}$ by conduction electrons of the metal substrate (Fig.~\ref{fig:odd_ring}d). Importantly, these two Kondo orbitals exhibit distinct spatial localization distributed over different parts of the \HuSR. To recover the experimentally observed spatially homogeneous distribution of the STS maps, simulated STS maps therefore need to include the contribution from Kondo orbitals associated with both doublets. Excellent agreement between experimental and simulated STS maps (Figs.~\ref{fig:odd_ring}a,b) supports the assignment to Kondo screening. Moreover, the delocalized Kondo signal over the entire molecule points out the frustrated magnetic structure of the synthesized odd-membered \HuSR\ as a consequence of the degenerate doublet ground state, as schematically depicted in Fig.~\ref{fig:odd_ring}c. In addition, Fig. S7 and Supplementary Fig. S22 show the natural transition orbitals for the allowed spin excitations between the degenerate doublet ground states and the excited quartet states. The experimentally observed homogeneous STS map at energies corresponding to the step-like features (e.g., \HuSR$^7$) is reproduced only when all symmetry-allowed $\Psi^D_{0,1}\to\Psi^Q_{2,3}$ transitions are taken into account (Fig.~\ref{fig:odd_ring}c). These results provide evidence for frustrated magnetic order arising from a degenerate ground state~\cite{Schnack2010,Kahn1997}.

\section*{Conclusion}

These results demonstrate that aromaticity can be leveraged as a design principle to control quantum states in $\pi$-magnetic carbon-based spin rings, enabling the creation of cyclic systems whose magnetic properties arise from long-range electronic correlations rather than local exchange. This bottom-up approach enables the design of diverse magnetic states with tunable spin excitations and spin frustration. This approach provides a foundation for engineering $\pi$-magnetic carbon-based molecular materials with strongly entangled electronic structures, with potential application in spintronics and quantum information technologies.

\section*{Methods}

\textbf{Precursor synthesis.} Molecular precursor (BCE-H$_3$-Tr) was synthesized from 4,6-dibromo-1H-phenalen-1-one by solution methods (see the detailed procedure and data in Supplementary Information).

\textbf{Sample preparation and STM/nc-AFM measurements.} A commercially available low-temperature Unisoku Joule-Thomson scanning probe microscope (1.2 K) and Liquid-helium free scanning probe microscope from CASAcme (4.3 K) operated at ultrahigh vacuum (3×10$^{-10}$ mbar) was used for all sample preparation and characterization. The Au(111) single crystals were cleaned using repeated cycles of Ar$^+$ sputtering and subsequent annealing to 950 K to obtain atomically flat terraces. The cleanliness of the crystals was checked by scanning with the STM before molecular deposition. Precursor BCE-H$_3$-Tr were thermally deposited on the clean Au(111) surface at 378 K for 1 hour. Precursor BCE-H$_3$-Tr was evaporated on the surface from a quartz crucible and the sublimation temperature was approximately 333 K. Then, the sample was transferred to a cryogenic scanner at 1.2 K for characterization. Carbon monoxide molecules were dosed onto the cold sample around 9 K (1×10$^{-8}$ mbar, 1 minute). A lock-in amplifier (531 Hz, 0.1-1 mV modulation) has been used to obtain dI/dV spectra. The STM and STS measurements were taken at 1.2 K, and the data were processed with the WSxM software. The bond-resolved nc-AFM images were acquired with Unisoku Joule-Thomson system operated at approximately 1.2 K. To achieve ultra-high spatial resolution, CO molecule was picked up from the Au(111) surface to the apex of a tungsten tip. A quartz tuning fork with a resonant frequency of 26 kHz has been used in nc-AFM measurements. The sensor was operated in frequency modulation mode with a constant oscillation amplitude of 0.3 Å. AFM measurements were performed in constant-height mode with Bias = 1 mV. STS mappings were acquired using Liquid-helium free scanning probe microscope. 

\textbf{CASCI calculations.} Molecular geometries were optimized in the high-spin state using density functional theory (DFT) as implemented in FHI-aims ~\cite{blum2009ab}. The PBE0 exchange--correlation functional~\cite{adamo1999toward} was used, and van der Waals interactions were included via the Tkatchenko–– Scheffler method~\cite{Tkatchenko2012}. Given the open-shell, multiradical character of the studied molecules, complete active space configuration interaction (CASCI) calculations were performed to accurately describe the wavefunctions and electronic energies~\cite{Kumar2025}. One- and two-electron integrals were constructed in the basis of molecular orbitals around the Fermi energy using the ORCA quantum chemistry package~\cite{neese2012orca}, with orbitals obtained from restricted open-shell Kohn-Sham (ROKS) calculations.  Active spaces of CAS(11,11) (odd-electron systems) and CAS(12,12) (even-electron systems) were employed, except for \HuSR$^{13}$, for which CAS(13,13) was used. The number of unpaired electrons was estimated from natural-orbital occupations obtained by diagonalizing the one-particle density matrix of the ground-state CASCI wavefunction.

\textbf{Kondo orbitals (KO).} Kondo orbitals were obtained by diagonalizing the Hamiltonian derived from the multi-channel Anderson model, which incorporates the many-body multiplet structure of molecules obtained from the CASCI neutral ground state and virtual charge states, as described in Ref.~\cite{calvo2024theoretical}.

\textbf{Natural transition orbitals (NTOs).} Natural transition orbitals (NTOs)~\cite{martin2003natural} were computed for the spin-flip process between the ground state and the first excited state to capture the spatial variation of the $dI/dV$ maps corresponding to IETS spin excitation maps.


Simulated nc-AFM and $dI/dV$ images were obtained using the probe-particle code~\cite{Hapala2014,Krej2017} with optimized geometries from total-energy DFT calculations.

\textbf{Radical character evaluation.} Radical character was evaluated using the approach proposed by Yamaguchi \,\textit{et al.}~\cite{yamaguchi1988extended}. Successive pairs of natural orbitals, consisting of the highest occupied natural orbital (HONO$-i$) and the lowest unoccupied natural orbital (LUNO$+i$), were considered. For each pair, a diradical coefficient $Y_d^{(i)}$ is defined as
\begin{equation}
Y_d^{(i)} = 1 - \frac{2 T_i}{1 + T_i^2},
\label{eq:yamaguchi_Yd}
\end{equation}
where
\begin{equation}
T_i = \frac{n_{\text{HONO}-i} - n_{\text{LUNO}+i}}{2}.
\label{eq:yamaguchi_T}
\end{equation}
Here, $n_{\text{HONO}-i}$ and $n_{\text{LUNO}+i}$ are the occupation numbers of the corresponding natural orbitals, and $i$ labels successive orbital pairs.

The total radical character is estimated by summing contributions from multiple orbital pairs,
\begin{equation}
Y_d^{\text{total}} = \sum_i Y_d^{(i)},
\end{equation}
which provides an effective measure of polyradical character when multiple orbital pairs exhibit substantial fractional occupation. For odd-membered rings, 0.5 is added to $Y_d^{\text{total}}$ to account for the SOMO contribution (one natural orbital with occupation 1.0).

\section*{Acknowledgments}

 We acknowledge the support from NSFC(22325203, 92577203, 22272050), GACR 25-17866X and the CzechNanoLab Research Infrastructure supported by MEYS CR (LM2023051). We also acknowledge the financial support from the TERAFIT project (CZ.02.01.01/00/22\_008/0004594) and Shanghai Jiao Tong University 2030 Initiative for financial support.

\section*{Competing Interests}

The authors declare no competing interests.

\section*{Author Contributions}

S.W., P.J.  and D.Y.L. conceived and supervised the project. P.J. and M.K. provided theoretical concept of H{\"u}ckel spint ring. M.K., D.S.P., E.M., L.V. and P.J. performed theoretical calculations and underlying interpretation. Z.Y., C.L. and Y.W. performed the STM/STS and nc-AFM measurements under the supervision of J.J., D.Y.L. and S.W.. Y.J.W. and X.Y.Z. synthesized the precursor molecules under the supervision of D.Y.L. and P.N.L.. P.J., M.K., D.Y.L. and S.W. wrote the manuscript with the input of all authors. All authors have given approval to the final version of the manuscript.

\clearpage

\bibliography{bib}
\bibliographystyle{naturemag}

\clearpage


\section*{SM}

\setcounter{figure}{0}
\renewcommand{\figurename}{ }
\renewcommand{\thefigure}{Fig}

\begin{figure}[!htb]
    \centering
    \includegraphics[width=1\linewidth]{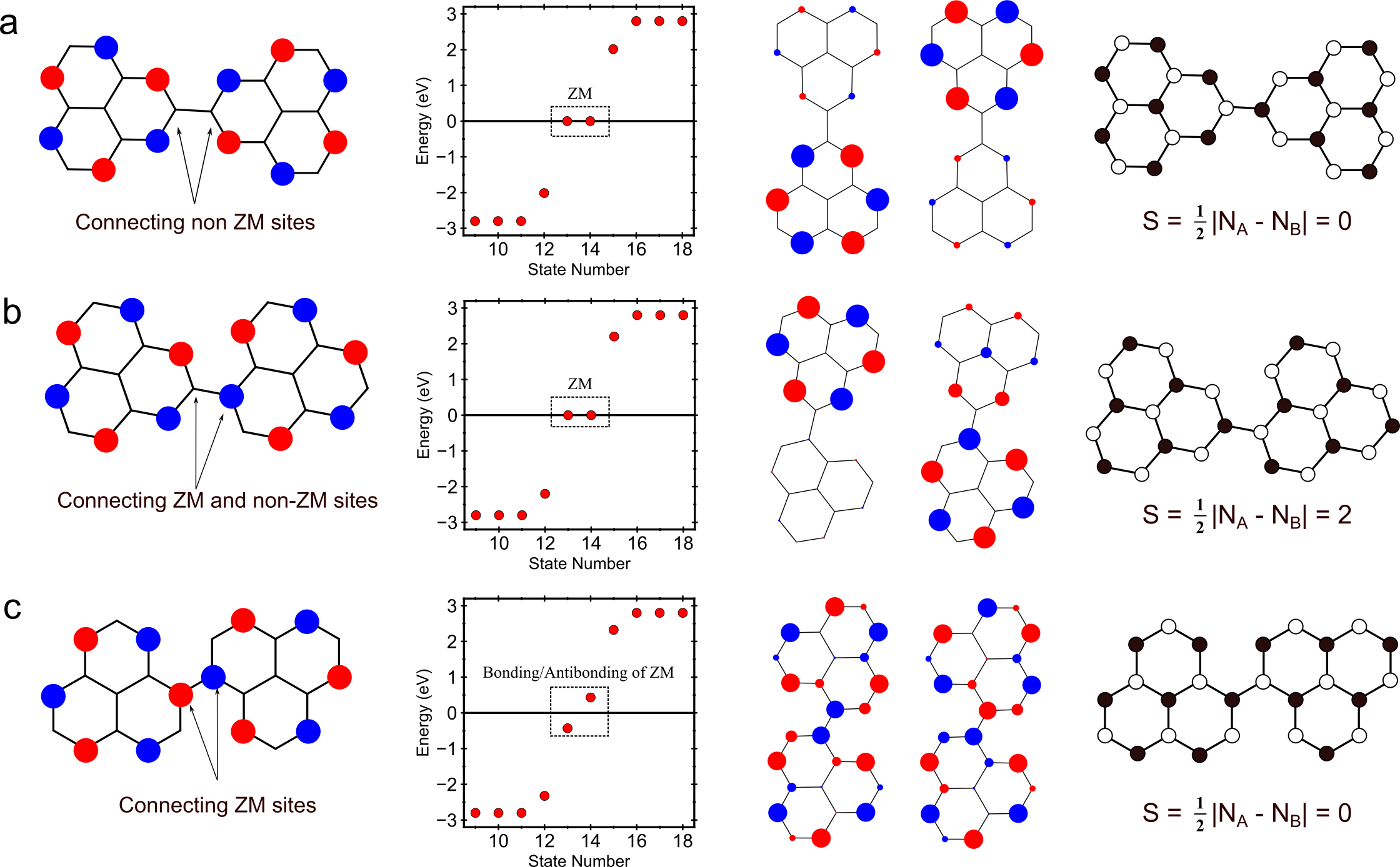}
    \caption{\textbf{S1 | H{\"u}ckel spectra for different connections in a [2]triangulene dimer.} Three distinct connection motifs between two [2]triangulene units, together with their corresponding H{\"u}ckel spectra (two orbitals around zero energy) and sublattice imbalance: \textbf{a}, connection between non--zero-mode (non-ZM) sites, \textbf{b}, connection between zero-mode (ZM) sites, and \textbf{c}, connection between a zero-mode (ZM) site and a non-ZM site.}
    \label{fig:ext_dimer_connection}
\end{figure}

\begin{figure}[!htb]
    \centering
    \includegraphics[width=1\linewidth]{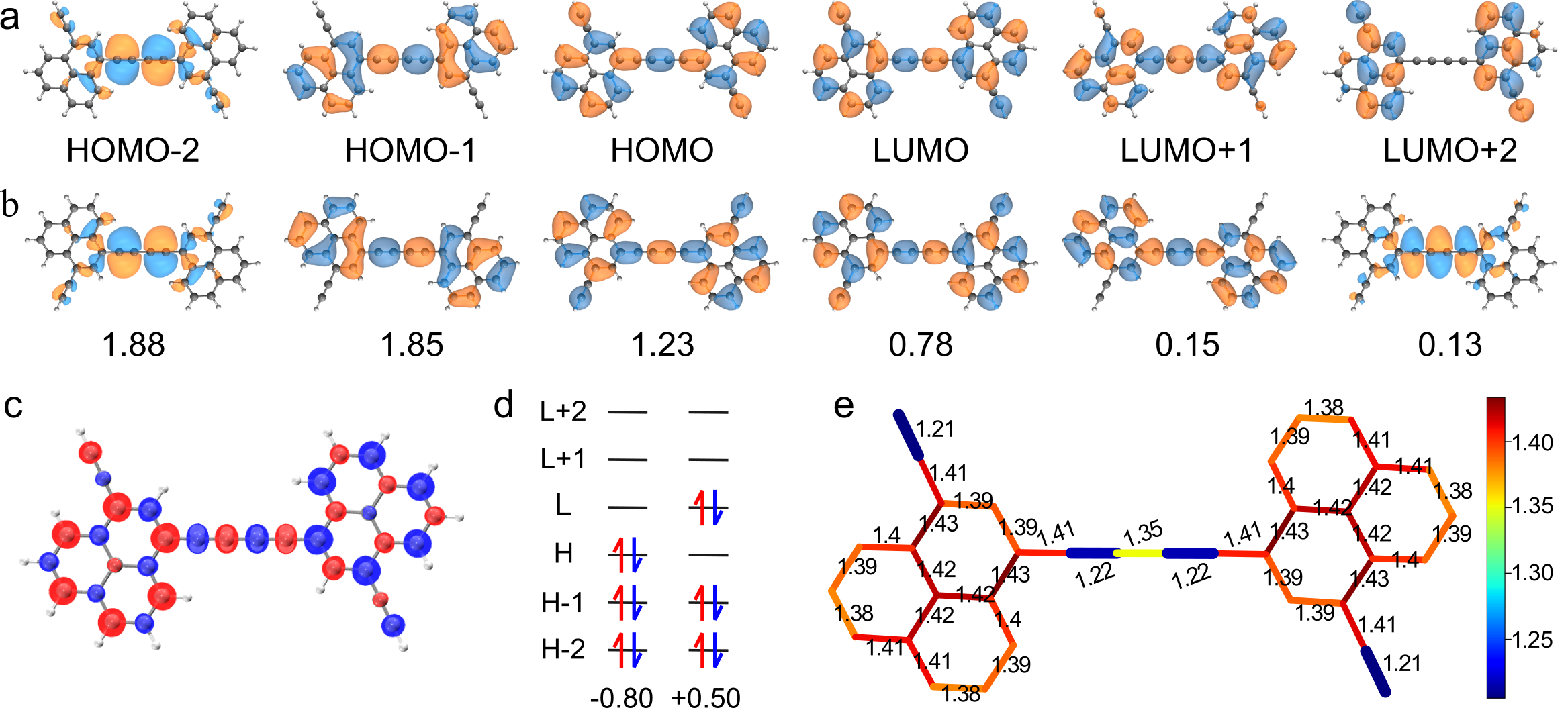}
    \caption{\textbf{S2 | Electronic structure of a [2]triangulene dimer.} \textbf{a}, Frontier DFT-PBE orbitals used for the DMRG calculation. \textbf{b}, Natural orbitals and their occupations obtained from DMRG. \textbf{c}, Isosurface of the spin density. \textbf{d}, Multireference wavefunction of the ground state. \textbf{e}, Bond lengths of the phenalene dimer are shown as a color map.}
    \label{fig:ext_dimer_nat}
\end{figure}

\begin{figure}[!htb]
    \centering
    \includegraphics[width=1\linewidth]{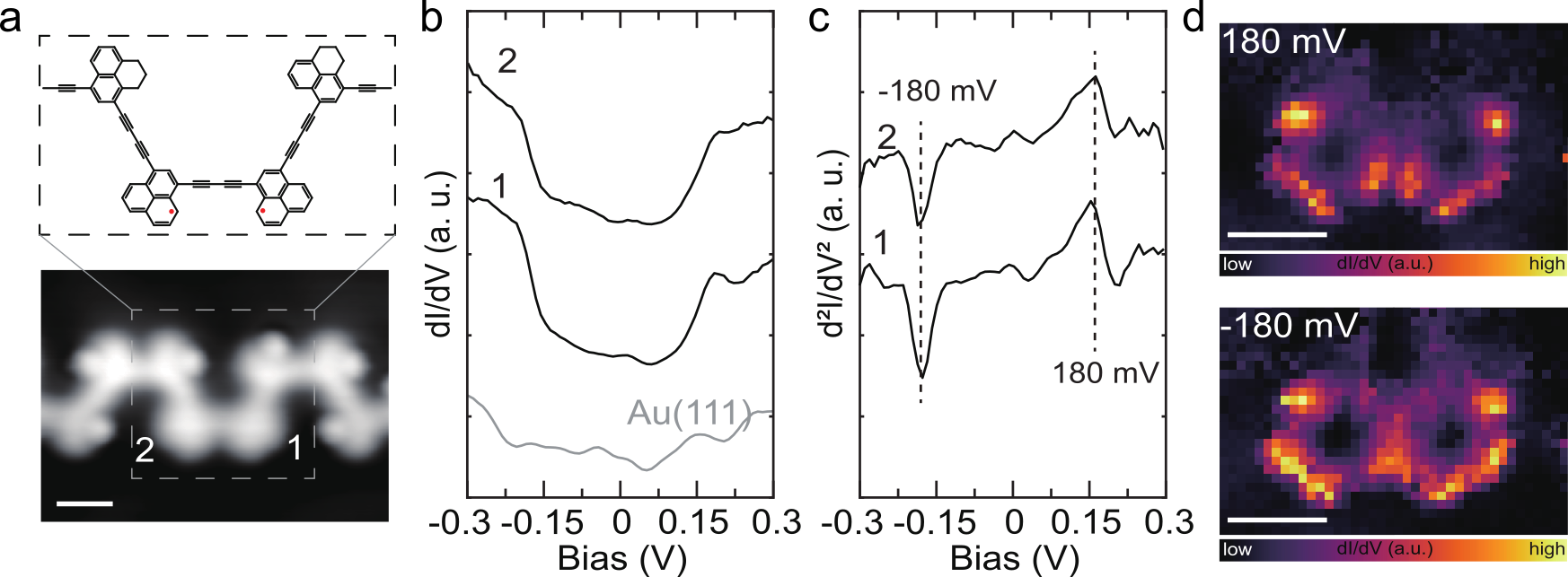}
    \caption{\textbf{S3 | Spin excitation of a spin-1/2 dimer embedded in a closed-shell long chain.} \textbf{a}, Chemical structure and STM images (300 mV, 10 pA) of a chain containing a spin-1/2 dimer; \textbf{b}, $dI/dV$ and \textbf{c}, $d^2I/dV^2$ spectra acquired on the activated spin sites; and \textbf{d}, STS maps acquired at 180 meV and $-180$ meV. Scale bar: 1 nm.}
    \label{fig:ext_dimer_exp}
\end{figure}

\begin{figure}[!htb]
    \centering
    \includegraphics[width=1\linewidth]{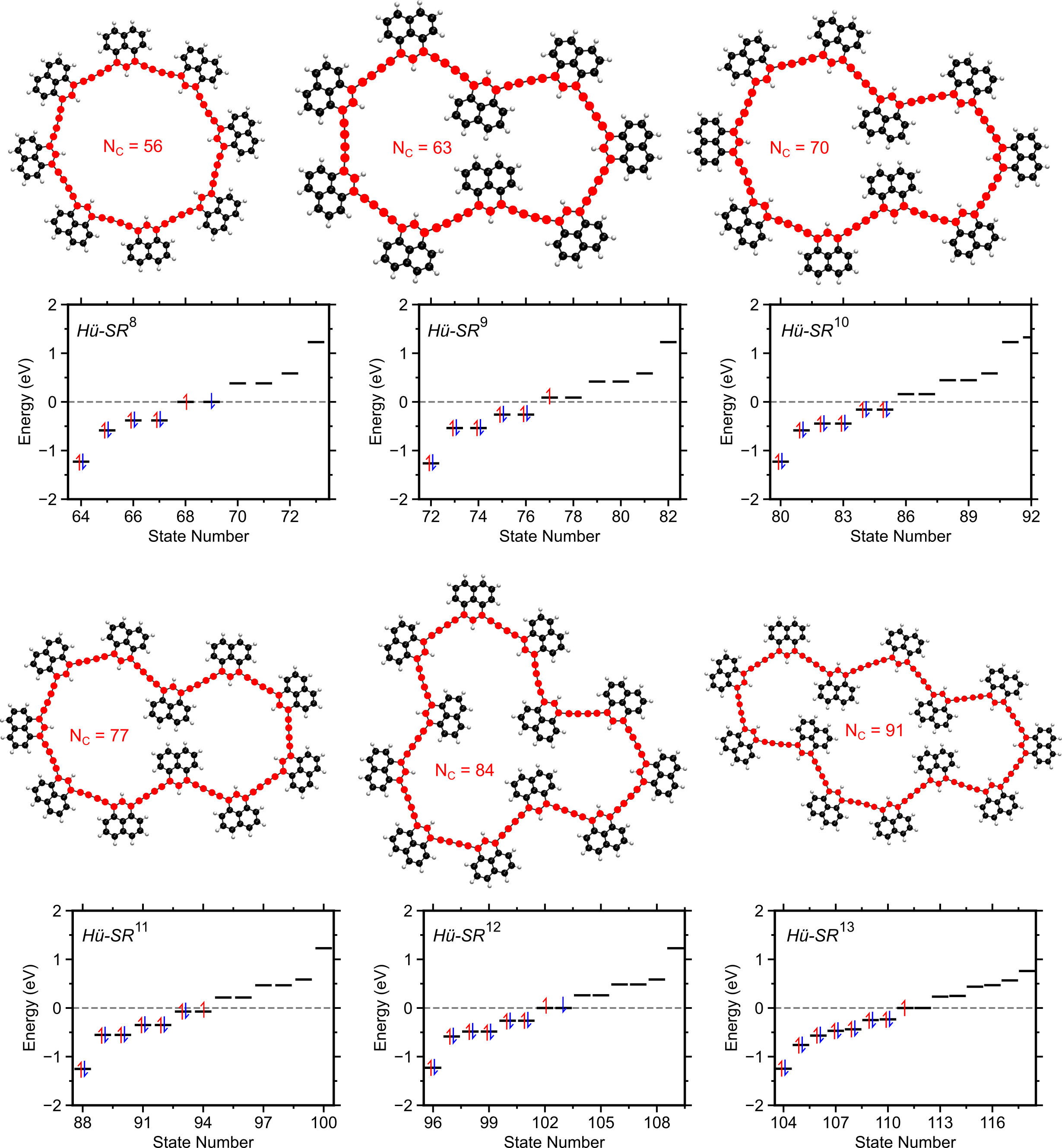}
    \caption{\textbf{S4 |} One-electron H{\"u}ckel energy spectra for spin rings in which [2]triangulene units are bridged by polyynic C$_4$ linkers (\HuSR$^8$ to \HuSR$^{13}$).}
    \label{fig:huckel_all}
\end{figure}

\begin{figure}[!htb]
    \centering
    \includegraphics[width=1\linewidth]{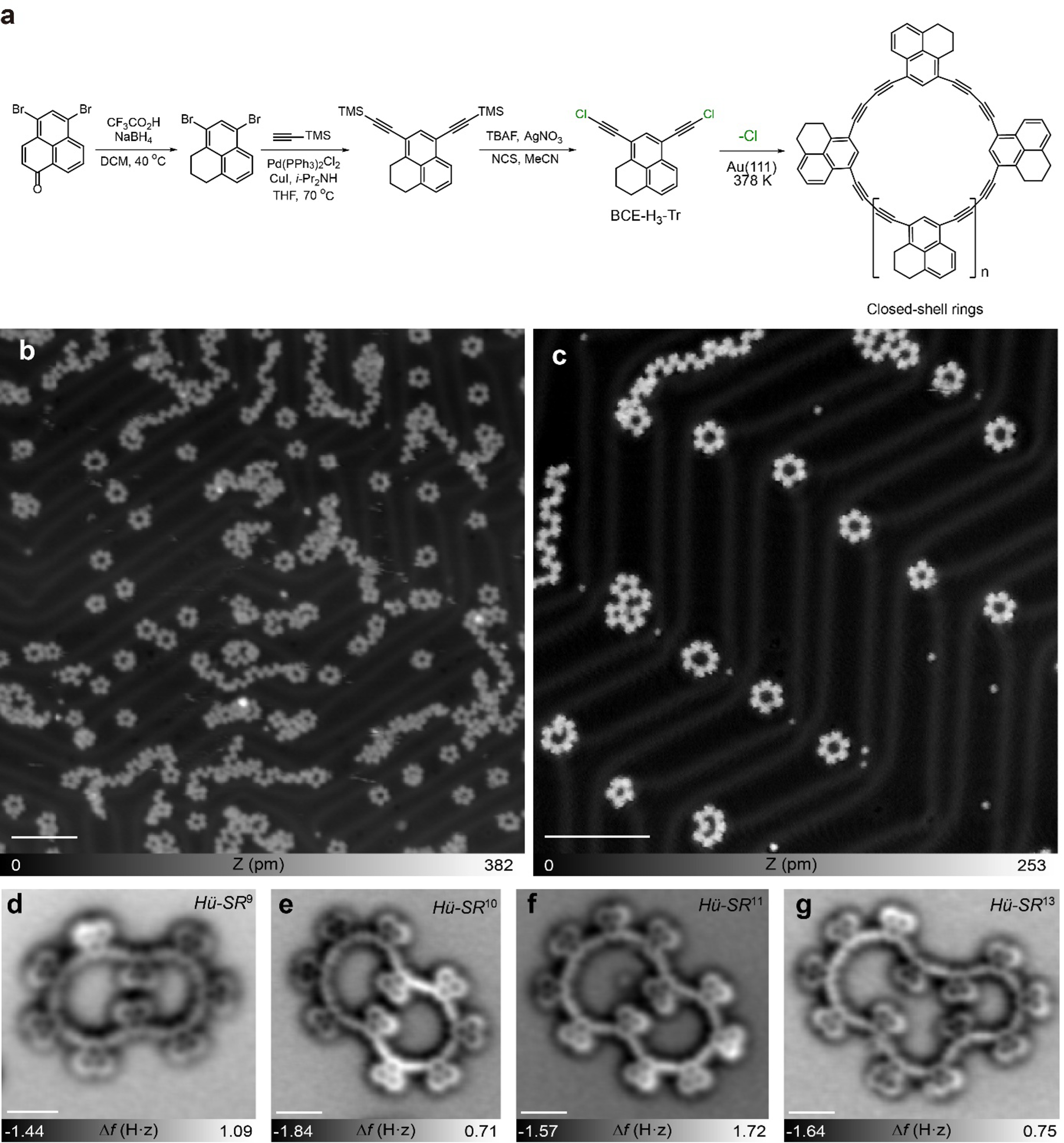}
    \caption{\textbf{S5 | Synthesis of molecular rings.} \textbf{a}, Synthesis scheme combining solution-phase chemistry and on-surface synthesis. \textbf{b,c}, Large-scale and zoomed-in STM images of closed-shell rings formed by depositing BCE-H$_3$-Tr on Au(111) and annealing at 378 K for 1 h (setpoints: $I=10$ pA and $V_\mathrm{bias}=1$ V in \textbf{b}; $I=10$ pA and $V_\mathrm{bias}=200$ mV in \textbf{c}; scale bars: 10 nm). \textbf{d--g}, nc-AFM images of larger H{\"u}ckel spin rings (\HuSR$^9$, \HuSR$^{10}$, \HuSR$^{11}$, and \HuSR$^{13}$) (setpoint: $V_\mathrm{bias}=1$ mV; scale bar: 1 nm).}
    \label{fig:ext_overview}
\end{figure}

\begin{figure}[!htb]
    \centering
    \includegraphics[width=1\linewidth]{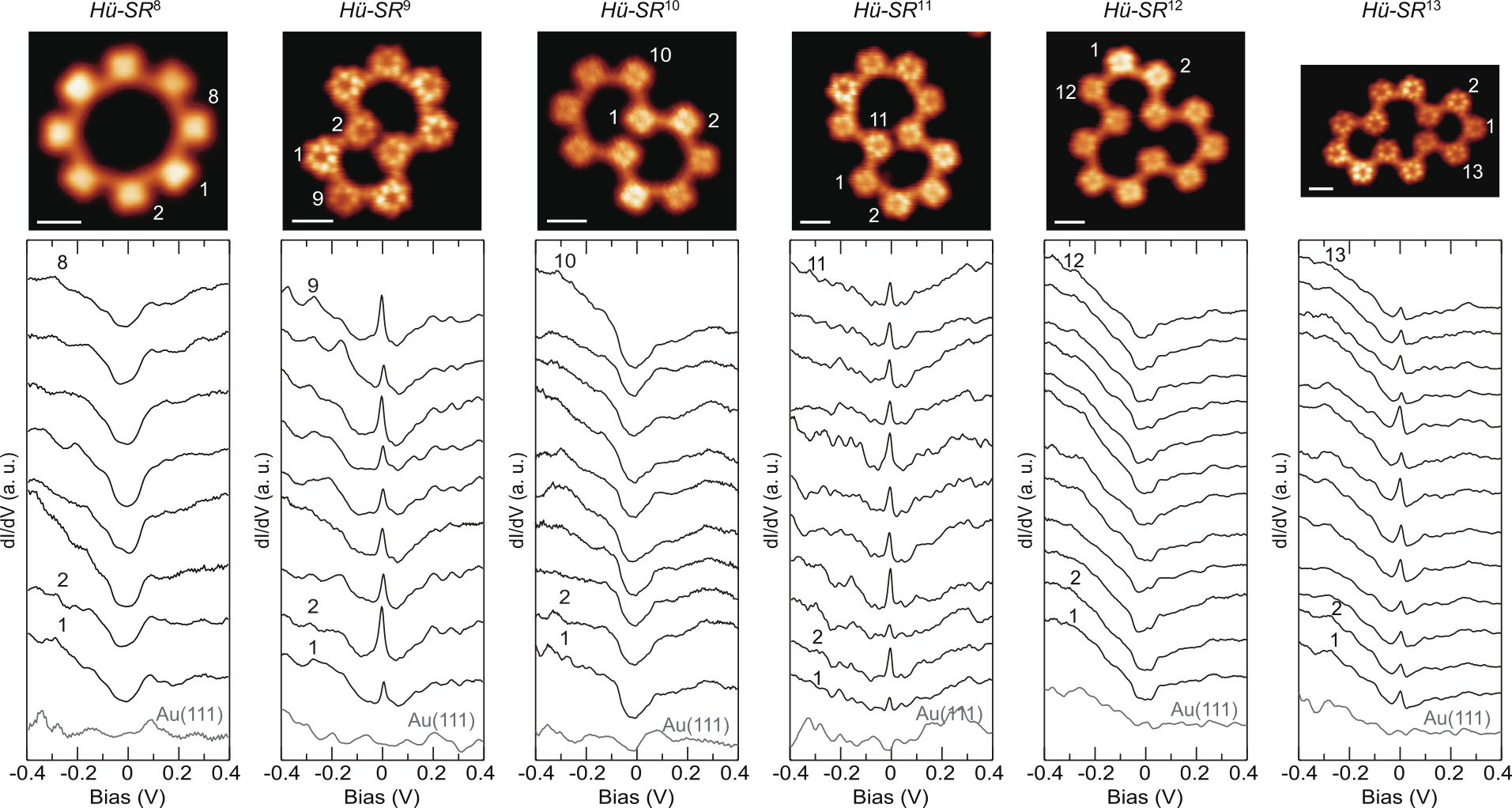}
\caption{\textbf{S6 | Characterization of larger molecular H{\"u}ckel spin rings.} Constant-height current images of \HuSR$^{8}$, \HuSR$^{9}$, \HuSR$^{10}$, \HuSR$^{11}$, \HuSR$^{12}$ and \HuSR$^{13}$, together with the corresponding $dI/dV$ spectra acquired on the activated spin sites (setpoints: $V_\mathrm{bias}=10$ mV for \HuSR$^{8}$, \HuSR$^{9}$, \HuSR$^{11}$ and \HuSR$^{13}$; $V_\mathrm{bias}=80$ mV for \HuSR$^{10}$; $V_\mathrm{bias}=140$ mV for \HuSR$^{12}$; scale bar: 1 nm).}
    \label{fig:ext_STS_long_ring}
\end{figure}

\begin{figure}[!htb]
    \centering
    \includegraphics[width=1\linewidth]{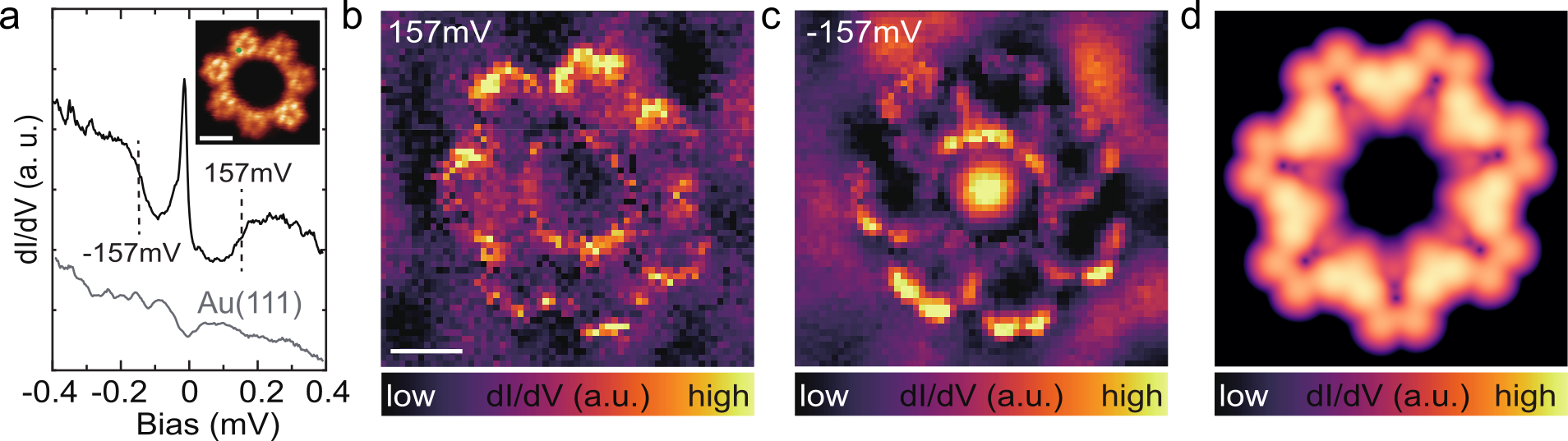}
\caption{\textbf{S7 | Spin excitation of \HuSR$^7$.} \textbf{a}, $dI/dV$ spectra of a Hü-SR7 (inset: constant-height current image, Bias = 10 mV) and \textbf{b,c}, the corresponding STS maps acquired at 157 mV and  -157 mV, respectively. \textbf{d}, Simulated STS maps computed from CASCI NTOs of \HuSR$^7$. Scale bars: 1 nm.}
    \label{fig:ext_STS_long_ring}
\end{figure}

\end{document}